# Discovery of terahertz-frequency orbitally-coupled magnons in a kagome ferromagnet

Mengqian Che[1], Weizhao Chen[2], Maoyuan Wang[3], F. Michael Bartram[1], Liangyang Liu[1], Xuebin Dong[4], Jinjin Liu[5,6], Yidian Li[1], Hao Lin[1], Zhiwei Wang[5,6], Enke Liu[4], Yugui Yao[5,6], Zhe Yuan[7,8], Guang-Ming Zhang[1,9,*], Luyi Yang[1,9,*]

[1]State Key Laboratory of Low Dimensional Quantum Physics and Department of Physics, Tsinghua University, Beijing, 100084, China
[2]Center for Advanced Quantum Studies and Department of Physics, Beijing Normal University, Beijing 100875, China
[3]Department of Physics, Xiamen University, Xiamen 361005, China
[4]Beijing National Center for Condensed Matter Physics and Institute of Physics, Chinese Academy of Sciences, Beijing 100190, China
[5]Centre for Quantum Physics, Key Laboratory of Advanced Optoelectronic Quantum Architecture and Measurement (MOE), School of Physics, Beijing Institute of Technology, Beijing, 100081, China
[6]Beijing Key Lab of Nanophotonics & Ultrafine Optoelectronic Systems, School of Physics, Beijing Institute of Technology, Beijing, 100081, China
[7]Institute for Nanoelectronic Devices and Quantum Computing, Fudan University, Shanghai 200433, China
[8]Interdisciplinary Center for Theoretical Physics and Information Sciences (ICTPIS), Fudan University, Shanghai 200433, China
[9]Frontier Science Center for Quantum Information, Beijing 100084, China
*Email: gmzhang@tsinghua.edu.cn; luyi-yang@mail.tsinghua.edu.cn

## Abstract

**In ferromagnetic materials, magnons - quanta of spin waves - typically resonate in the gigahertz range. Beyond conventional magnons, while theoretical studies have predicted magnons associated with orbital magnetic moments, their direct observation has remained challenging. Here, we present the discovery of two distinct terahertz orbitally-coupled magnon resonances in the topological kagome ferromagnet $Co_3Sn_2S_2$. Using time-resolved Kerr rotation spectroscopy, we pinpoint two magnon resonances at 0.61 and 0.49 THz at 6 K, surpassing all previously reported magnon resonances in ferromagnets due to strong magnetocrystalline anisotropy. These dual modes originate from the strong coupling of localized spin and orbital magnetic moments. These findings unveil a novel category of magnons stemming from orbital magnetic moments, and position $Co_3Sn_2S_2$ as a promising candidate for high-speed terahertz spintronic applications.**

Magnons are the elementary quanta of low-energy collective spin excitations in magnetically ordered materials. These excitations can be exploited for information processing, transport and storage [1, 2]. Magnonics, an emerging field centered around such uses, has the potential to replace today's CMOS (complementary metal oxide semiconductor) technology for more compact, lower dissipation and faster computation devices. Magnonic devices based on conventional ferromagnetic materials suffer from limited operation speed, which is determined by the ferromagnetic resonance (FMR) frequency, typically in the gigahertz range [3, 4]. While antiferromagnets have much higher resonance frequencies (up to terahertz [1, 5]), they are very difficult to control and detect due to their vanishing net magnetization. To realize ultrafast computation beyond today's GHz clock rate, it is crucial to discover novel magnetic materials that are both easily controllable and resonant in the sub-terahertz to terahertz frequency regime.

In addition to conventional magnons, recent theoretical studies have hinted at the existence of unconventional magnons arising from orbital magnetic moments [6-12]. However, orbital magnetic moments are usually much smaller than spin magnetic moments because of the quenching of the orbital angular momentum. Consequently, directly observing magnons associated with orbital magnetic moments presents a significant experimental challenge. Magnetic kagome lattice materials, such as $Co_3Sn_2S_2$, $Fe_3Sn_2$, FeSn and $Mn_3Sn$ [13-17], offer the potential to realize such exotic orbitally-coupled magnon states. This unique lattice structure creates an intricate interplay between geometry, magnetism and topology, giving rise to a plethora of exotic quantum phenomena, including Weyl nodes, frustrated magnetic states, flat-band correlations and unconventional superconductivity [18]. In particular, the emergence of flat bands induced by the kagome lattice leads to pronounced electron localization and orbital magnetic moments in real space. Experimental investigations into these novel orbital magnetic moments and their associated orbital magnon excitations are still at an early stage.

$Co_3Sn_2S_2$ is a pioneering magnetic Weyl semimetal and a star player in this captivating world. The coupling between its electronic wavefunction topology and magnetic spin configuration gives rise to many intriguing properties, including a colossal anomalous Hall effect [13, 19], a giant anomalous Nernst effect [20], and a pronounced magneto-optical response [21]. Cobalt atoms in $Co_3Sn_2S_2$ are arranged on a kagome lattice (Fig. 1A). This material favors magnetization perpendicular to its kagome plane due to a strong easy-axis magnetocrystalline anisotropy [13, 22, 23]. This strong anisotropy could lead to the generation of high-frequency magnons, which as mentioned previously are desirable for magnonic applications. Yet, direct observation of these elusive magnons and their ultrafast dynamics remains an appealing quest. Furthermore, scanning tunneling microscopy studies have revealed orbital magnetic moments associated with the flat bands of the surface kagome layer [24] and localized spin-orbit polarons surrounding sulfur vacancies [25, 26], making further investigation of magnons involving orbital magnetic moments in this material and their interaction with spin magnetic moments a compelling avenue for further research. Ultrafast optical spectroscopy holds the key to not only directly probing high-frequency magnons and their ultrafast dynamics but also understanding the precise nature of this enigmatic system.

Here we investigate ultrafast coherent magnon dynamics in single crystals of $Co_3Sn_2S_2$ using time-resolved magneto-optical Kerr effect (trMOKE) spectroscopy. To our surprise, we directly observed two magnon modes in the terahertz range (0.61 and 0.49 THz at 6 K) in the time domain. This marks the highest magnon resonance frequencies ever reported in ferromagnets. The exceptionally high magnon frequencies are attributed to the strong magnetocrystalline anisotropy and exchange anisotropy. Using the standard Kittel model [27], we find that the extracted Landé g-factors deviate from the anticipated value of 2, suggesting an interplay between spin and orbital magnetic moments. Supported by a microscopic model, we propose that these dual modes emerge from the low-energy collective excitations of coupled spin and orbital magnetic moments in the ferromagnetic ordered state. Both spin and orbital ordered magnetic moments can be determined experimentally, qualitatively consistent with the density functional theory (DFT) calculations [24]. Therefore, our work uncovers a novel type of magnons due to orbital magnetic moments and lays the foundation for the development of terahertz spintronic devices using topological kagome ferromagnets.

## Record high dual terahertz ferromagnetic resonances

To study the ultrafast magnon dynamics, we used a two-color trMOKE setup [28]. Both pump and probe beams were at normal incidence (along the easy axis of the $Co_3Sn_2S_2$ crystal (z-direction)). An external magnetic field was applied at a small angle (about 7 degrees) with the sample plane. We performed measurements on three samples from two crystal growth laboratories (IOP and BIT), and found consistent results across all samples.

Figure 1B illustrates a possible mechanism for coherent magnon generation and detection through trMOKE. Initially, the magnetic moment (**M**) aligns with the effective magnetic field (**H**$_{eff}$), which is determined by both the out-of-plane anisotropy field and the near in-plane external field. At *t* = 0, upon the arrival of the pump pulse, mechanisms such as the reduction of the out-of-plane anisotropy field due to laser heating as depicted in the schematic induce a sudden change in the direction of the effective field. As a result, the new effective field direction (**H'**$_{eff}$) is no longer parallel with **M**, generating a torque that initiates subsequent precession of **M** about **H'**$_{eff}$ at the FMR frequency. The Gilbert damping causes **M** to spiral towards **H'**$_{eff}$, leading to the damped oscillations of the magnetization. The magneto-optical Kerr effect allows changes in the z-component of the magnetization ($\Delta M_z(t)$) to be detected by the time-delayed probe pulse.

Figure 1C shows the raw trMOKE signal (black curve) acquired at 7 T and 6 K. The signal can be decomposed into two distinct parts: an exponential decay (yellow curve) and an oscillatory decay (blue curve). The exponential decay is attributed to energy transfer from the pump beam to the coupled electron-lattice-spin system, inducing ultrafast demagnetization within a picosecond (at low temperatures) followed by a gradual recovery at longer timescales [29-31]. Our present work only focuses on the oscillatory decay signal due to coherent magnon precession.

To obtain the magnetization precession signal, we subtracted the exponential decay background from the raw data, and then used Fast Fourier Transform (FFT) analysis to identify the precession frequencies. As depicted in Fig. 1D, the FFT clearly shows two distinct resonances: a dominant one (Mode H) at a remarkably high frequency of ~0.60 THz and a weaker resonance (Mode L) at a slightly lower frequency of ~0.48 THz at 7 T (the frequencies approximate 0.61 THz and 0.49 THz when extrapolated to zero field, see Fig. 2). Both resonance frequencies exceed typical FMR frequencies by 1-2 orders of magnitude [3, 4], making them, to the best of our knowledge, the highest magnetic resonance frequencies ever reported in ferromagnets.

Intriguingly, the observation of two magnon modes is highly unusual. Given that there are three Co atoms per unit cell in the kagome plane, we anticipate three magnon branches: one acoustic mode where all three magnetic moments precess in phase, and two optical modes with out-of-phase precession of the magnetic moments (Fig. 1E). The in-phase precession of the acoustic mode sums the amplitudes across inequivalent Co sublattices, while the out-of-phase precession of the optical modes should completely cancel out any net change in magnetization, $\Delta M_z$. Therefore, only the coherent acoustic mode is detectable in our experiment. Experimentally, Mode H is the dominant mode, corresponding to a spin wave gap of ~2.5 meV, in good agreement with inelastic neutron scattering measurements [32, 33]. The presence of a weaker magnon resonance indicates the existence of a second type of magnons, likely linked to orbital magnetic moments of the flat band (discussed later), which is also a coherent acoustic mode.

Figure 1F displays the trMOKE signal at a variety of external magnetic fields at 6 K. The oscillatory signal clearly grows with increasing field strength. To quantify this behavior, we extracted the oscillation amplitude of Mode H and plotted it as a function of the applied field in Fig. 1G (the amplitude of Mode L is too small at low temperatures and low fields). This amplitude can be modeled by assuming the coherent magnon generation mechanism shown in Fig. 1B. Critically, this magnon generation mechanism aligns with the previously mentioned acoustic mode, because the pump excitation initializes all Co moments with the same phase. Notably, at higher temperatures where the amplitude of Mode L becomes more prominent, its field dependence mirrors that of Mode H. This shared scaling behavior suggests that Mode L is also a coherent acoustic mode.

## Magnetic field and temperature dependence

Figure 2 visualizes the full field dependence of the oscillatory modes using FFT analysis (after subtracting the exponential decay background). It plots the FFT amplitude as a function of both field and frequency at various temperatures, revealing the existence of two distinct temperature regimes. Below 149 K, both modes are clearly distinguishable and exhibit a redshift with increasing field and temperature. Notably, their frequency separation shrinks as the temperature rises. Above 149 K, only one resonance can be resolved due to broader linewidths relative to the frequency separation. At 152 – 155 K, the singly-resolved frequency switches from redshift to blueshift with increasing field (the fitted central frequency is shown as orange dots). At even higher temperatures, the resonance blueshifts linearly with the applied field, indicating that the precession is primarily driven by the external field.

The dependence of magnon frequencies on the applied field was first analyzed by the Kittel equation of FMR resonance [27]:

$$\omega = \gamma\{[H_{\text{ext}}\cos(\theta_0 - \theta_H) + H_{\text{ani}}(\cos\theta_0)^2][H_{\text{ext}}\cos(\theta_0 - \theta_H) + H_{\text{ani}}\cos(2\theta_0)]\}^{\frac{1}{2}}, \quad (1)$$

where $\gamma = \frac{g\mu_B}{\hbar}$ is the electron gyromagnetic ratio, $g$ is the Landé g-factor, $\mu_B$ is the Bohr magneton, $H_{\text{ani}}$ is an effective internal magnetic field, corresponding to the combined magnetocrystalline anisotropy field and exchange anisotropy field, $H_{\text{ext}}$ is the external field, $\theta_H$ is the angle between the sample normal and the applied field, and $\theta_0$ is the angle between the equilibrium magnetization direction and the sample normal, which is determined by minimizing the free energy:

$$H_{\text{ani}}\sin\theta_0\cos\theta_0 + H_{\text{ext}}\sin(\theta_0 - \theta_H) = 0. \quad (2)$$

As illustrated in the bottom-right panel of Fig. 2, the Kittel model shows a transition in frequency from a redshift to blueshift with increasing applied field when $H_{\text{ext}} \sim H_{\text{ani}}$.

Both individual magnon modes below 150 K and the combined resonance above that temperature fit with the Kittel equation remarkably well. The data from three samples consistently overlap. The similar $H_{\text{ani}}$ values across the two modes suggest that both modes experience comparable internal anisotropy fields. We further exploited this similarity by performing a global fit and assuming the same $H_{\text{ani}}$ for both modes. The extracted parameters are shown in Fig. 3A,B. These global fits yield smaller error bars compared to the individual fits, demonstrating the improved accuracy of this shared-$H_{\text{ani}}$ approach.

Our measurements reveal an exceptionally large anisotropy field: $20.5 \pm 0.1$ T at 4 K, in line with the magnetization measurements [22] and inelastic neutron scattering experiments [32]. Remarkably, this anisotropy field surpasses those found in other layered structures, such as $Fe_3GeTe_2$ with 5 T [34], $Cr_2Ge_2Te_6$ with 0.5 T [35], $Fe_3Sn_2$ with 1.1 T [36] and $CrI_3$ with 3 T [37]. The strong anisotropy in $Co_3Sn_2S_2$ arises from the potent interaction between electron spins and orbital motion (spin-orbit coupling), which is a key feature of materials with interesting band topology. With increasing temperature, $H_{\text{ani}}$ steadily decreases, ultimately vanishing around the Curie temperature (160-165 K). Notably, this measured transition temperature is 10 – 15 K lower than the reported value due to laser heating effects, which can be estimated by a model in Ref. [38], using the thermal conductivity data [39] and heat capacity data [40].

Interestingly, Fig. 3B shows that the g-factors of the two modes exhibit striking differences: Mode H has a g-factor >2 and Mode L has a g-factor < 2 (consistent across all fitting routines and samples). Notably, both g-factors remain relatively temperature-independent below 150 K, where the modes are distinguishable.

Using global (individual) fitting, the averages shown as dashed lines are 2.19±0.06 (2.21±0.08) for Mode H and 1.81±0.04 (1.59±0.21) for Mode L. As the temperature rises from 150 to 165 K, where the modes merge, the g-factor fluctuates around 2 before abruptly dropping to ~1.3 at ~165 K, coinciding with the vanishing of $H_{ani}$.

Figure 2 also reveals the amplitude of Mode L increases (relative to Mode H) with temperature, approaching that of Mode H at ~149 K. This behavior is further dissected through simultaneously fitting both real and imaginary FFT components with two Lorentzian functions (or with two exponentially decaying cosines in the time domain), allowing precise extraction of each mode's amplitude. Figure 3C showcases a set of FFT real part data with the corresponding fits at various temperatures at 7 T (the imaginary parts omitted for clarity). Crucially, the extracted amplitude ratio of Mode H to Mode L steadily decreases from ~5 to ~1 as the temperature increases from 6 to 140 K (Fig. 3D). This trend suggests a rising magnetization in Mode L relative to Mode H with temperature. In contrast, at a given temperature, the amplitude ratio exhibits weak dependence on the magnetic field (Fig. 3E).

## Origin of the two magnon resonances

Contrary to conventional expectations, our experiment has revealed two distinct coherent acoustic magnon resonances, each characterized by unique g-factors which significantly deviate from the anticipated spin g-factor of 2. This surprising finding suggests that the observed magnon modes likely arise from the interplay between spin and orbital magnetic moments, facilitated by the strong spin-orbit coupling of $Co_3Sn_2S_2$. To elucidate these findings, we propose that the dual magnon modes stem from the low-energy collective excitations of coupled local spin and orbital magnetic moments in the ferromagnetic ordered state. In kagome magnets, it has been suggested that, in addition to localized spin moments, there also exists localized orbital magnetism [41]. As illustrated in Fig. 4A, the highly anisotropic *d*-orbitals in $Co_3Sn_2S_2$ tend to organize themselves in the hexagonal rings of the kagome lattice. This arrangement confines the *d*-electron orbital motion within each hexagon due to the perfect cancellation of their propagating wavefunctions on the kagome lattice [41]. These orbital moments, coupled to the spin magnetic moments, contribute to the formation of a long-range ferromagnetic order. The low-energy collective excitations of the coupled localized spin and orbital magnetic moments give rise to the two coupled magnon modes.

In order to confirm this picture, we propose an effective microscopic model to capture the essential physics: a coupled spin and orbital magnetic system. The energy of the system per unit cell can be expressed as:

$$U = -\mu_S \left( J \sum_{<ij>} \hat{S}_i \cdot \hat{S}_j + \frac{1}{2} H_{KS} \sum_i (\hat{S}_i \cdot \hat{z})^2 + \sum_i \boldsymbol{H}_{\text{ext}} \cdot \hat{S}_i \right) \\ - \mu_L \left( Q \sum_{<ij>} \hat{L}_i \cdot \hat{L}_j + \frac{1}{2} H_{KL} \sum_i (\hat{L}_i \cdot \hat{z})^2 - \sum_i \boldsymbol{H}_{\text{ext}} \cdot \hat{L}_i \right) - \frac{\lambda}{\mu_B} \mu_S \mu_L \sum_i \hat{S}_i \cdot \hat{L}_i, \quad (3)$$

where <ij> means summation over the nearest neighbor lattice sites, and the directions of each spin and orbital magnetic moment are represented by a unit vector $\hat{S}_i$ and $\hat{L}_i$ with magnitudes denoted by $\mu_S$ and $\mu_L$, respectively. $J, Q > 0$ are isotropic ferromagnetic Heisenberg exchange coupling, chosen for its simplicity (extending this model to include anisotropic exchange interactions is straightforward). $H_{KS}$ and $H_{KL}$ are the magnetocrystalline anisotropic fields, $H_{\text{ext}}$ is the external field, and $\lambda$ describes the coupling between spin and orbital magnetic moments. Notably, the Zeeman energy for the orbital part reaches a minimum when the orbital magnetic moment is antiparallel to the applied field. Such an alignment accounts for the experimental observations of the diamagnetic nature of the orbital magnetic moment [24, 26].

Then the equilibrium positions of $\hat{S}_i$ and $\hat{L}_i$ are determined by minimizing $U$. Magnon frequencies are calculated using the (undamped) Landau-Lifshitz-Gilbert (LLG) equation, where the effective magnetic field is given by $\boldsymbol{H}_{\eta,i}^{\text{eff}} = -\frac{\partial U}{\partial \boldsymbol{\mu}_{\eta,i}}$ and the gyromagnetic ratio $\gamma_\eta = \frac{\mu_B}{\hbar} g_\eta$ with $\eta = S$ or $L$, and $g_S = 2$ and $g_L = 1$. The equations are linearized around the equilibrium at each field value to obtain the eigenmodes and their corresponding frequencies. In this coupled spin and orbital system, there are two acoustic magnon modes corresponding to Mode H and Mode L, along with four optical magnon modes. Notably, the frequencies of the acoustic modes are independent of $J$ and $Q$, but depend on all other parameters; the optical modes are not detectable in our experiment due to a zero net magnetization change.

Following numerical fitting, we observe a perfect alignment between our spin-orbital coupled model and the experimental data, as evidenced by the fits depicted in Fig. 4B. The corresponding fitting parameters for all samples are displayed in Fig. 4C. At low temperatures, the spin magnetic moment is close to $0.3\mu_B$, consistent with previous neutron scattering and magnetization experiments [13, 42], and it gradually decreases as the temperature rises, matching magnetization measurements [22, 33, 43]. The orbital magnetic moment is determined to be around $0.007\mu_B$ at low temperatures, in good agreement with DFT calculations predicting $0.003\mu_B$/Co [24], and it also diminishes with increasing temperature. Notably, the spin-orbit coupling parameter is insensitive to temperature changes, maintaining a value around 45 – 50 T. Furthermore, the anisotropy field mirrors the behavior plotted in Fig. 3A obtained using the Kittel model.

Given the small orbital magnetic moment of $0.007\mu_B$, it is still puzzling how the second mode is observed in our experiment. To unravel this mystery, we further examine the eigenvectors of the coupled LLG equations. As shown in Fig. 4D (simulation of the 60 K data), the coupling between the spin and orbital magnetic moments causes both Mode H and Mode L to exhibit spin and orbit components, with the eigenvector polar angle components denoted by $d\theta_S^M$ and $d\theta_L^M$, respectively, where the superscript $M$ represents Mode H or Mode L. The measured Kerr signal for each mode ($\Delta\theta_{\text{Kerr}}^M$) is proportional to $\mu_S \sin(\theta_S) \delta\theta_S d\theta_S^M + a\mu_L \sin(\theta_L) \delta\theta_L d\theta_L^M$, where $a$ represents the Kerr rotation sensitivity difference between spin and orbital magnetic moments and is taken as 1 for simplicity; $\sin(\theta_{S/L})$ characterizes the projection of the magnetic moment change on the measurement axis, where $\theta_{S/L}$ signifies the equilibrium position; $\delta\theta_{S/L}$ denotes the initial offset of the equilibrium position with respect to ***H'***eff due to the pump excitation. Then we calculated the Kerr rotation amplitude ratio of Mode H over Mode L ($\Delta\theta_{\text{Kerr}}^{\text{Mode H}}/\Delta\theta_{\text{Kerr}}^{\text{Mode L}}$) as a function of magnetic field and temperature, respectively, shown in Fig. 4E,F. Remarkably, this calculated ratio aligns closely with our experimental data shown in Fig. 3E,D. As the temperature rises, the spin magnetic moment has an increasing weight in Mode L (*i.e.*, $d\theta_S^{\text{Mode L}}$ increases), making it more pronounced in the Kerr signal.

Moreover, we would like to emphasize that the determined $\mu_L$ alone is exceedingly small. In the absence of spin-orbit coupling, orbital magnons are decoupled from spin magnons, rendering them exceptionally challenging to be detected experimentally. In $Co_3Sn_2S_2$, nevertheless, the strong spin-orbit coupling significantly mixes the spin and orbital motions. Since $\mu_S \gg \mu_L$, the measured signals for both modes are predominantly influenced by the spin magnon component. Consequently, two spin-orbit-coupled magnon modes manifest experimentally. Furthermore, the surprisingly high FMR frequencies in $Co_3Sn_2S_2$ also enable the clear observation of these two well-separated coupled modes. In contrast, conventional ferromagnets have much lower resonance frequencies, resulting in frequency separations that could be too small to be resolved within the linewidths.

In summary, we have systematically studied the coherent magnon dynamics in the topological kagome ferromagnet $Co_3Sn_2S_2$ using trMOKE, revealing two distinct terahertz-frequency spin-orbital-coupled magnon modes. These extremely high-frequency coupled magnon modes arise from the strong spin-orbit coupling inherent in the topological kagome ferromagnet. Our findings not only set a new record for magnon frequencies in ferromagnetic materials but also experimentally identify a novel class of magnons associated with orbital magnetic moments. This research provides valuable insights into the interplay between magnetism and topology in a magnetic topological material and opens up promising avenues for the development of innovative terahertz spintronic devices utilizing topological kagome magnets.

## Acknowledgements

We thank Wanjun Jiang for helpful discussions. This work was supported by the National Key R&D Program of China (Grant Nos. 2020YFA0308800 and 2021YFA1400100) and the National Natural Science Foundation of China (Grant Nos. 12361141826 and 12074212). G.M.Z. acknowledges the support of National Key Research and Development Program of China (Grant No. 2023YFA1406400). Y.Y. acknowledges the support of the National Natural Science Foundation of China (Grant No. 12321004).

## Author contributions:

L.Y. conceived and supervised the project.  M.C. built the time-resolved experiment and performed the measurements with help from F.M.B., L.L., Y.L, and H.L.  M.C. carried out all the data analysis.  X.D., J.L. Z.W. and E.L. grew the samples.  G.M.Z. proposed the coupled spin-orbital model and outlined the related physical picture to the experiments.  W.C., M.W., Y.Y. and Z.Y. provided theoretical insights.  M.C. and L.Y. wrote the paper with the help of G.M.Z and input from all coauthors.

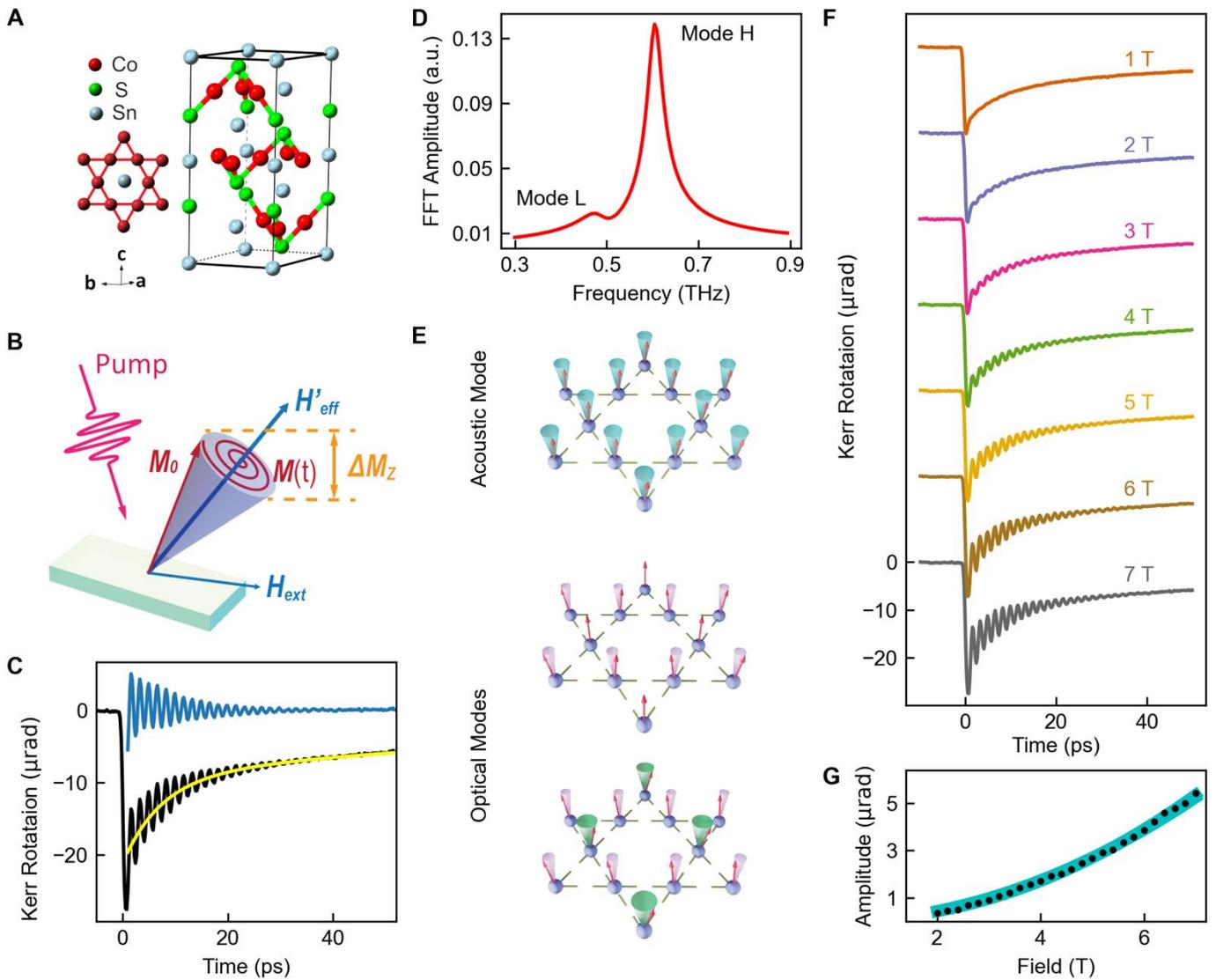

**Fig. 1. Magnon modes revealed by time-resolved Kerr rotation measurements with in-plane fields at 6 K.**
(**A**) Crystal structure of $Co_3Sn_2S_2$. (**B**) Schematic of the coherent magnon generation mechanism. (**C**) trMOKE signal at 7 T. The raw data (black curve) consists of an exponential decay (yellow curve) and an oscillatory decay. Subtracting the exponential decaying background allows the magnon oscillation to be seen more clearly (blue curve), where multiple frequencies are visible. (**D**) Fast Fourier Transform (FFT) analysis of the subtracted data showing two distinct magnon modes. a.u. stands for arbitrary units. (**E**) Schematic of three eigenmodes of magnons. The top one is an acoustic (in-phase) mode, which is observable in our experiment. The others are optical (out-of-phase) modes, resulting in zero net magnetization change (undetectable). (**F**) trMOKE signals with different external fields, showing increasing oscillatory signal with the applied field. Curves are offset vertically for clarity. (**G**) Oscillation amplitude of Mode H versus field (black dots). The cyan line represents a fit.

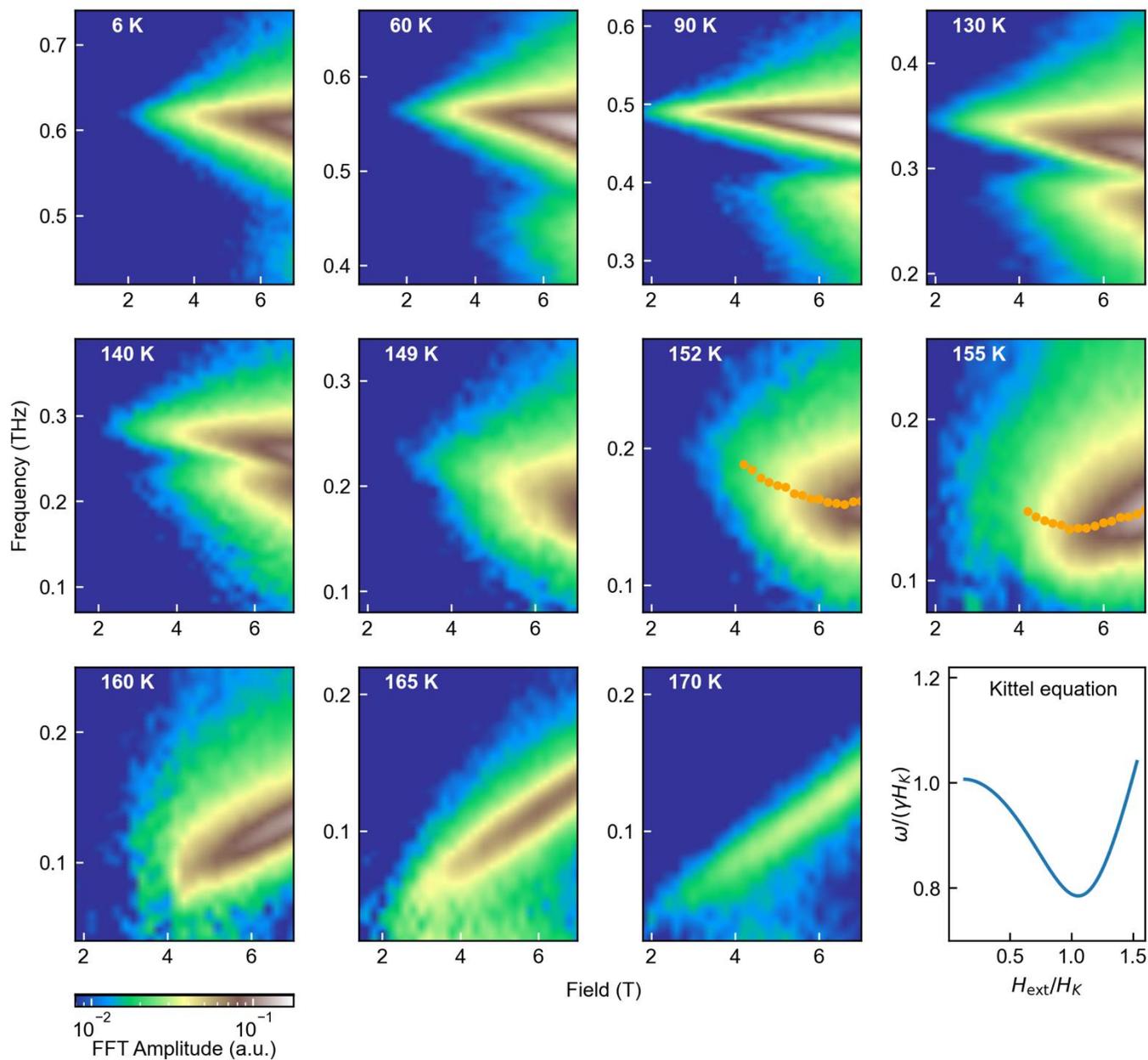

**Fig. 2. Field- and temperature-dependence of magnons.** FFT amplitude of time-resolved Kerr rotation signals as a function of frequency and applied magnetic field at various temperatures. Two modes are distinguishable below ~150 K, but merge together above that temperature due to decreased frequency separation with respect to the linewidths. At low (high) temperatures, the resonant frequencies decrease (increase) with increasing field. At 152 and 155 K, the central frequency (orange dots) turns from blueshift to redshift with the applied field, indicating the internal field becomes comparable with the applied field. The bottom-right panel shows the Kittel equation of normalized frequency versus normalized applied external field. a.u. stands for arbitrary units.

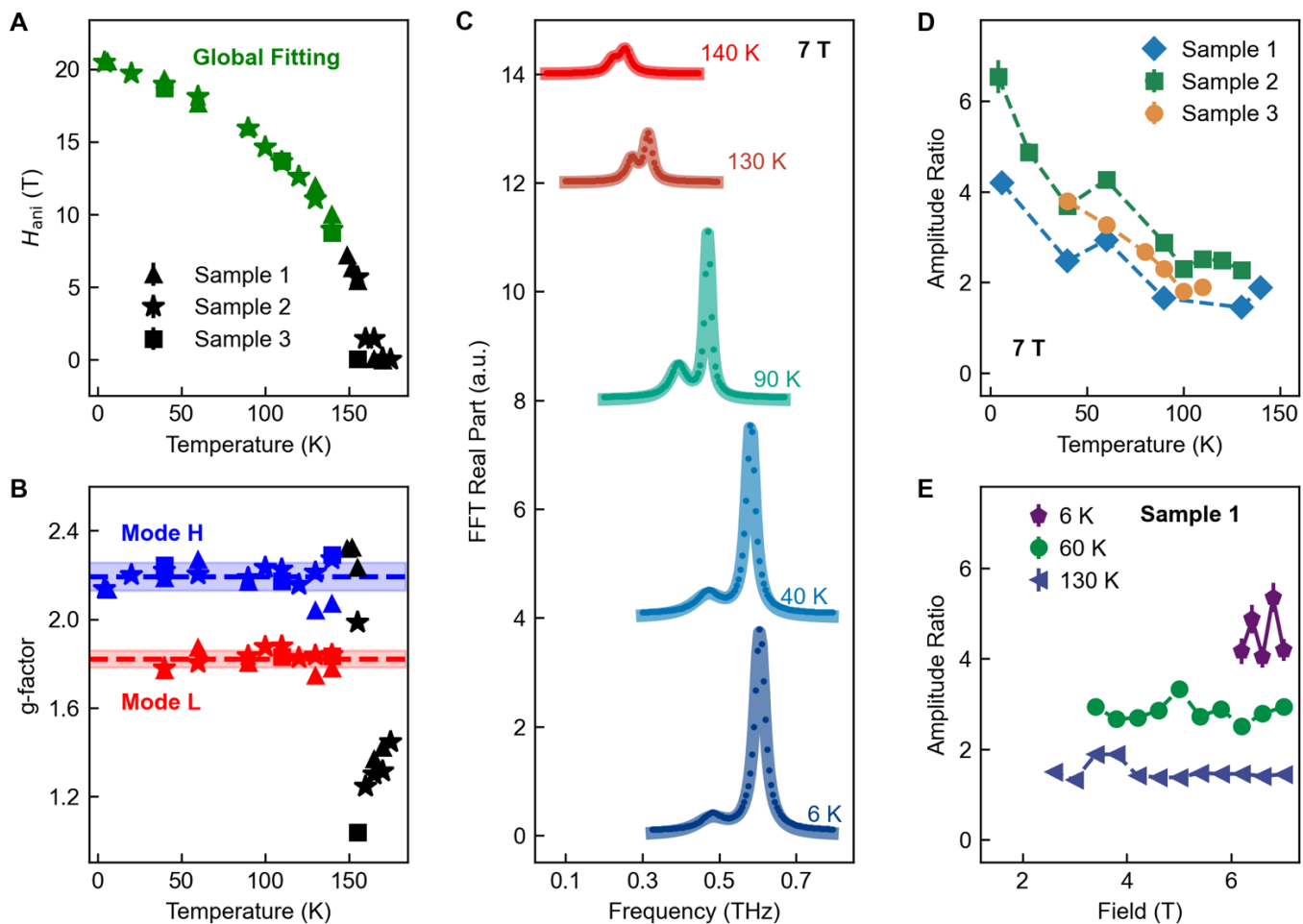

**Fig. 3. Anisotropy field ($H_{\mathrm{ani}}$), Landé g-factor and magnon amplitude ratio of Mode H over Mode L.** Anisotropy field $H_{\mathrm{ani}}$ (**A**) and Landé g-factor (**B**) were extracted using the Kittel equation (see text). At and above ~150 K, only one resonance is visible, the extracted data are shown in black. Below ~150 K, Modes H (blue) and L (red) were fit either individually (not shown) or globally (shown here) by assuming the same $H_{\mathrm{ani}}$. In either case, the two modes show distinct Landé g-factors. The dashed lines in (**B**) represent the average values of the g-factors, with the shaded areas indicating the associated uncertainties. (**C**) FFT real part at 7 T for different temperatures. a.u. stands for arbitrary units. The curves are offset vertically for clarity. The dots represent the raw data, and the lines show the corresponding fitted curves. As the temperature increases, there is a noticeable decrease in the ratio ($A_H/A_L$) of the extracted amplitudes (**D**). In contrast, this amplitude ratio exhibits minimal dependence on the field at a fixed temperature (**E**).

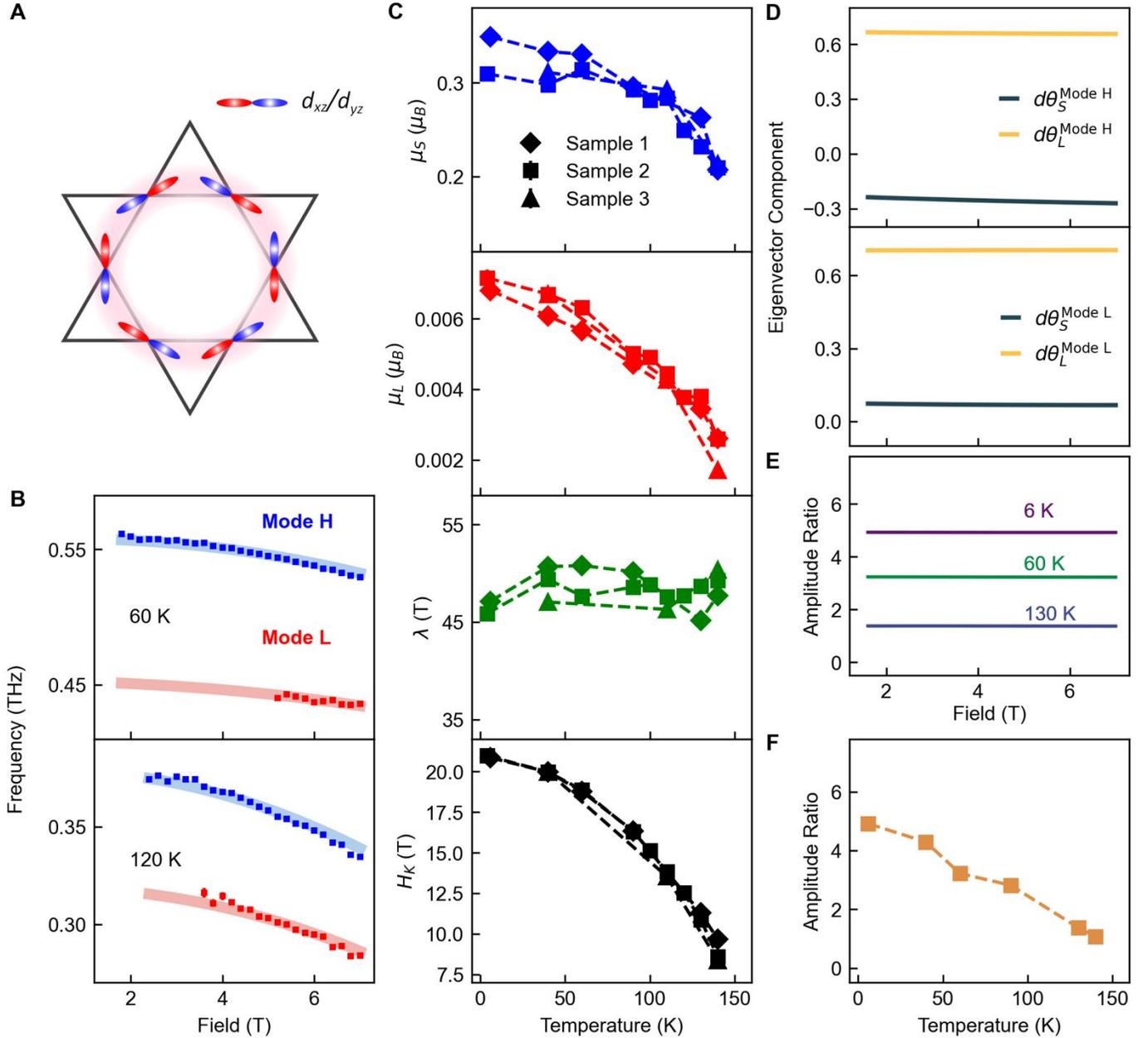

**Fig. 4. Effective model of coupled spin-orbital magnon modes.** (**A**) Top view of the orbital configuration ($d_{xz}/d_{yz}$) consisting of the flat band near the Fermi level. (**B**) Examples of fits at 60 and 120 K using the coupled spin and orbital magnon model (see text). Squares are the experimental data and solid lines are the fits. (**C**) Extracted fitting parameters of the spin magnetic moment ($\mu_S$), orbital magnetic moment ($\mu_L$), spin-orbit coupling parameter ($\lambda$) and anisotropy field ($H_K$) as a function of temperature for all samples. The magnetocrystalline anisotropic fields were chosen to be the same in order to reduce the fitting parameters (*i.e.*, $H_{KS} = H_{KL} = H_K$). (**D**) Eigenvector components, $d\theta_S^M$ and $d\theta_L^M$, where the superscript *M* represents Mode H or Mode L, both of which contain spin and orbital components. Calculated amplitude ratio of Mode H over Mode L is plotted as a function of field (**E**) and temperature (**F**), where the average fitting parameters (**C**) of different samples were used in the calculation.

16